\documentclass[12pt,preprint]{aastex}
\usepackage{psfig}

\slugcomment{KSUPT-06/2 \hspace{0.5truecm} February 2006}

\newcommand{\kmsmpc}{{\rm \, km\, s}^{-1}{\rm Mpc}^{-1}}

\begin{document}

\title{Supernova Ia and Galaxy Cluster Gas Mass Fraction Constraints on
Dark Energy}

\author{Kyle M. Wilson, Gang Chen\footnote{Present address: Institute for Astronomy, 2680 Woodlawn
Drive, Honolulu, HI 96822; gchen@ifa.hawa\-ii.edu}, and Bharat Ratra
}

\affil{Department of Physics, Kansas State University, 116 Cardwell
Hall, Manhattan, KS 66506; \mbox{kmwilson}, \mbox{chengang},
\mbox{ratra@phys.ksu.edu}}

\begin{abstract}
We use the Riess et al.\ (2004) supernova Ia apparent magnitude versus redshift data and the Allen et al.\ (2004) galaxy cluster gas mass
fraction versus redshift data to constrain dark energy models. These data provide complementary constraints that when combined together
significantly restrict model parameters and favor slowly-evolving dark energy density models, close to the Einstein cosmological constant
limit of dark energy.
\end{abstract}


\section{Introduction}

Less than a decade ago, astronomers discovered that the expansion of the Universe was speeding up (Riess et al.\ 1998; Perlmutter et al.\
1999).  In the context of Einstein's general relativity, this acceleration is attributed to dark energy that varies slowly with time and
space, if at all.  More recent supernova Ia (SNIa) redshift-magnitude data,\footnote{ For the SNIa data see, e.g., Riess et al.\ (2004,
hereafter R04), Nobili et al.\ (2005), Clocchiatti et al.\ (2006), and Astier et al.\ (2006). SNIa data constraints on cosmological
parameters are also discussed by Wang \& Tegmark (2005),  Lazkoz et al.\ (2005), Hannestad (2005), Shafieloo et al.\ (2006), Zhang \& Wu
(2005), God\l{}owski \& Szyd\l{}owski (2005), Sethi et al.\ (2005), Bento et al.\ (2005), and Jassal et al.\ (2006).} and cosmic microwave
background (CMB) anisotropy data --- that indicates the Universe is spatially flat (see, e.g., Podariu et al.\ 2001b; Durrer et al.\ 2003;
Melchiorri \& \"{O}dman 2003; Bennett et al.\ 2003; Page et al.\ 2003) --- combined with a low mean value for the nonrelativistic matter
density parameter $\Omega_M$ (Chen \& Ratra 2003b, and references therein), 
strengthens the claim that a previously
undetected dark energy accounts for a major part of the cosmic energy budget. For reviews of the current situation see Peebles \& Ratra
(2003), Steinhardt (2003), Carroll (2004), Padmanabhan (2005), and Perivolaropoulos (2006).

While CMB anisotropy and SNIa constraints on dark energy will continue to improve,\footnote{For near future SNIa constraints see Sollerman
et al.\ (2005) and Hamuy et al.\ (2006).  Within a decade or so, the proposed JDEM/SNAP space mission should provide even tighter
constraints (see http://snap.lbl.gov/ and Podariu et al.\ 2001a; Crotts et al.\ 2005; Albert et al.\ 2005, and references therein).
Mukherjee et al.\ (2003), Caldwell \& Doran (2004), and Lee et al.\ (2006) discuss CMB constraints on dark energy.} it is important to use
many different tests to constrain dark energy models.  This will allow for consistency checks, as well as provide complementary constraints
on dark energy.  A number of other tests have been considered.  These include the redshift-angular size test (see, e.g., Chen \& Ratra
2003a; Podariu et al.\ 2003;  Jackson 2004; Puetzfeld et al.\ 2005; Daly \& Djorgovski 2005), the strong gravitational lensing test (see,
e.g., Fukugita et al.\ 1990; Turner 1990; Ratra \& Quillen 1992; Chae et al.\ 2004; Sereno 2005; Alcaniz et al.\ 2005), the galaxy cluster
gas mass fraction versus redshift test (see, e.g., Sasaki 1996; Pen 1997; Allen et al.\ 2004, hereafter A04; Chen \& Ratra 2004, hereafter
CR; Rapetti et al.\ 2005; Alcaniz \& Zhu 2005; Kravtsov et al.\ 2005; Chang et al.\ 2006), and the baryon acoustic oscillation test (see,
e.g., Glazebrook \& Blake 2005; Angulo et al.\ 2005; Wang 2006). In addition, structure formation in dark energy models will soon provide a
useful diagnostic of model parameters (see, e.g., Horellou \& Berge 2005; McDonald et al.\ 2006; Le Delliou 2005; Bartelmann et al.\ 2005;
Percival 2005; Mainini 2005). In this paper we constrain dark energy 
models by doing a joint analysis of the  SNIa apparent magnitude
versus redshift data of R04 and the galaxy cluster gas mass fraction versus redshift data of A04.

There are a number of different dark energy models under current discussion.\footnote{ Recent papers include Neupane \& Wiltshire (2005),
Pietroni (2005), Zhao et al.\ (2005), Capozziello et al.\ (2005), Fran\c{c}a (2005), Barenboim et al.\ (2005), Barnes et al.\ (2005),
Abdel-Rahman \& Riad (2005), Barger et al.\ (2005), Sol\'{a} \& \v{S}tefan\v{c}i\'{c} (2006), and Arbey (2006).} The current ``standard"
cosmological model is $\Lambda$CDM (Peebles 1984), where the energy density of the low-redshift Universe is dominated by a cosmological
constant $\Lambda$ (with a constant energy density $\rho_\Lambda$), with nonrelativistic matter --- mostly cold dark matter (CDM) ---
playing a subdominant role. Models in which the dark energy is due to a slowly varying, in time and space, scalar field ($\phi$), have also
attracted recent attention.  A prototypical $\phi$CDM example has scalar field potential energy density $V(\phi) \propto \phi^{-\alpha},
\alpha>0$, at low redshift (Peebles \& Ratra 1988; Ratra \& Peebles 1988); again CDM plays a subdominant role.  Also discussed is the XCDM
parametrization for time-varying dark energy. This parametrization approximates dark energy by a fluid with a negative time-independent
equation of state parameter $w=P/\rho$, where $P$ is the fluid pressure and $\rho$ the energy density. This is an inaccurate approximation
during the scalar field dominated epoch when $w$ is time dependent (see, e.g., Ratra 1991).  We consider spatially flat spacetimes for the
$\phi$CDM model and the XCDM parametrization, but allow spatial curvature to be a free parameter in the $\Lambda$CDM case. In this paper we
constrain parameters of these three models by using SNIa and galaxy cluster data.

In $\S\: 2$ we show how we use the supernova and galaxy cluster data to constrain cosmological parameters.  Results are presented and
discussed in $\S \:3$. We conclude in $\S \:4$.

\section{Computation}

A04 list x-ray gas mass fractions for 26 rich clusters, determined from $Chandra$ observations. These clusters lie at redshifts between
0.08 and 0.89. Following A04 and CR, we use this data to determine the probability distribution function (likelihood) $L^G (\Omega_M, p,
h)$. Here $h$ is the Hubble constant in units of 100 $\kmsmpc$, and $p$ is the cosmological constant density parameter $\Omega_{\Lambda}$
for the $\Lambda$CDM model, $w$ for the XCDM parametrization, and $\alpha$ for the $\phi$CDM model. To derive $L^G (\Omega_M, p, h)$ we
marginalize over the bias factor $b$, as well as over $\Omega_bh^2$, where 
$\Omega_b$ is the baryonic mass density parameter.  Later we
will need to marginalize over $h$.  We account for uncertainties in $b$, $\Omega_bh^2$, and $h$ by using Gaussian priors with $b=0.824 \pm
0.089$, $\Omega_bh^2=0.0214 \pm 0.002$, and $h=0.72 \pm 0.08$, all one standard deviation errors.\footnote{ See Ganga et al.\ (1997) for a
discussion of this method. As noted in CR, there is a more precise estimate of $h$ than the value quoted above, $h=0.68 \pm 0.04$ (Gott et
al.\ 2001; Chen et al.\ 2003, and we have halved the quoted two standard deviation errors).  Also, the $\Omega_bh^2$ value used above is consistent 
with the estimate from the WMAP CMB anisotropy measurement and the 
primordial deuterium abundance measurement, but significantly higher than 
an estimate based on the primordial helium and lithium abundance 
measurements (see, e.g., Peebles \& Ratra 2003;
Steigman 2005; Fields \& Sarkar 2006), so we also consider the Peebles \& Ratra (2003) summary estimate $\Omega_bh^2=0.014 \pm 0.004$.
Constraints derived using the values of $h$ and $\Omega_bh^2$ quoted in 
this footnote are shown as darker dashed lines in Figs.\ 2---4, while
constraints from the priors quoted in the main text are shown using 
lighter continuous lines.} See A04 and CR for more detailed discussions of
the procedure we use.

We also use the SNIa apparent magnitude versus redshift data from R04, in particular, the gold data set of 156 SNIa with redshifts up to
almost 1.8. From this data we determine the probability distribution function $L^S (\Omega_M, p, h)$, following, e.g., R04 or Podariu \&
Ratra (2000).

The joint likelihood for the SNIa and galaxy cluster data is the product of the two individual likelihoods. The two-dimensional probability
distribution function for $\Omega_M$ and $p$, $L(\Omega_M, p)$, is determined by marginalizing this product over $h$. For each of the three
models mentioned above, we compute $L(\Omega_M, p)$ on the two-dimensional $(\Omega_M, p)$ grid.  The 1, 2, and 3 $\sigma$ confidence
contours are the set of points where the likelihood is $e^{-2.30/2}$, $e^{-6.17/2}$, and $e^{-11.8/2}$ of the maximum likelihood value.

\section{Results and Discussion}

Figure 1 shows the R04 gold SNIa constraints on the $\phi$CDM model with $V(\phi) \propto \phi^{-\alpha}$, determined by marginalizing
$L^S(\Omega_M, \alpha, h)$ over $h$. These contours are significantly more constraining than those derived by Podariu \& Ratra (2000, Fig.\ 1) from 
earlier SNIa data.

Figure 2 shows the combined SNIa and galaxy cluster constraints on the $\Lambda$CDM model.  These joint constraints are significantly
tighter than those derived using either the SNIa data (R04, Fig.\ 8) or the galaxy cluster data (A04, Figs.\ 4 and 8; CR, Fig.\ 1) alone.
This is because the galaxy cluster data tend to tightly constrain $\Omega_M$ (A04, CR), while the SNIa data tend to tightly constrain a
linear combination of $\Omega_\Lambda$ and $\Omega_M$ (R04). Together, the 
data focus attention on a small part of parameter space near $\Omega_M 
\sim 0.3$ and $\Omega_\Lambda \sim 0.7$ where the Universe is spatially 
flat.

Figure 3 shows the joint data constraints on the XCDM parametrization. Models close to $\Omega_M \sim 0.3$ and $w \sim -1$ are favored.  We
emphasize that $w=-1$ corresponds to the spatially-flat $\Lambda$CDM model.

Figure 4 shows the joint data constraints on the $\phi$CDM model with $V(\phi) \propto \phi^{-\alpha}$.  Models close to $\Omega_M \sim
0.25$ and $\alpha \sim 0$ are favored.  We note that $\alpha = 0$ corresponds to the spatially-flat $\Lambda$CDM model.

The SNIa and galaxy cluster data together favor a spatially-flat $\Lambda$CDM model with $\Omega_M \approx 0.25 - 0.3$.  It might be
significant that other data also favor this value of $\Omega_M$ (see, e.g., Chen \& Ratra 2003b; Spergel et al.\ 2003).  We emphasize,
however, that slowly-evolving dark energy models are also consistent with this data.

\section{Conclusion}

We use supernova Ia apparent magnitude and x-ray cluster gas mass fraction data to jointly constrain cosmological parameters.  These data
sets provide complementary constraints and together they tightly constrain cosmological parameters.  The spatially-flat $\Lambda$CDM model
is favored, although a $\phi$CDM model with slowly-decreasing dark energy cannot yet be ruled out.

More and better SNIa and galaxy cluster data will allow for tighter constraints on model parameters.  Equally important would be a
resolution of the conflict between the two different sets of estimates of $\Omega_b$, as well as tighter limits on $h$.  All these are
likely to happen in the next few years.

\bigskip

We acknowledge useful discussion with M.\ Sayler, K.\ Teramura, and J.\ Whitmer, and support from DOE grants DE-FG03-99ER41093 and
DE-FG02-00ER45824, NASA grants AISR NAG5-11996 and ATP NAG5-12101, and NSF grants AST-0206243, AST-0434413, and ITR 1120201-128440.

\begin{figure}[p] \psfig{file=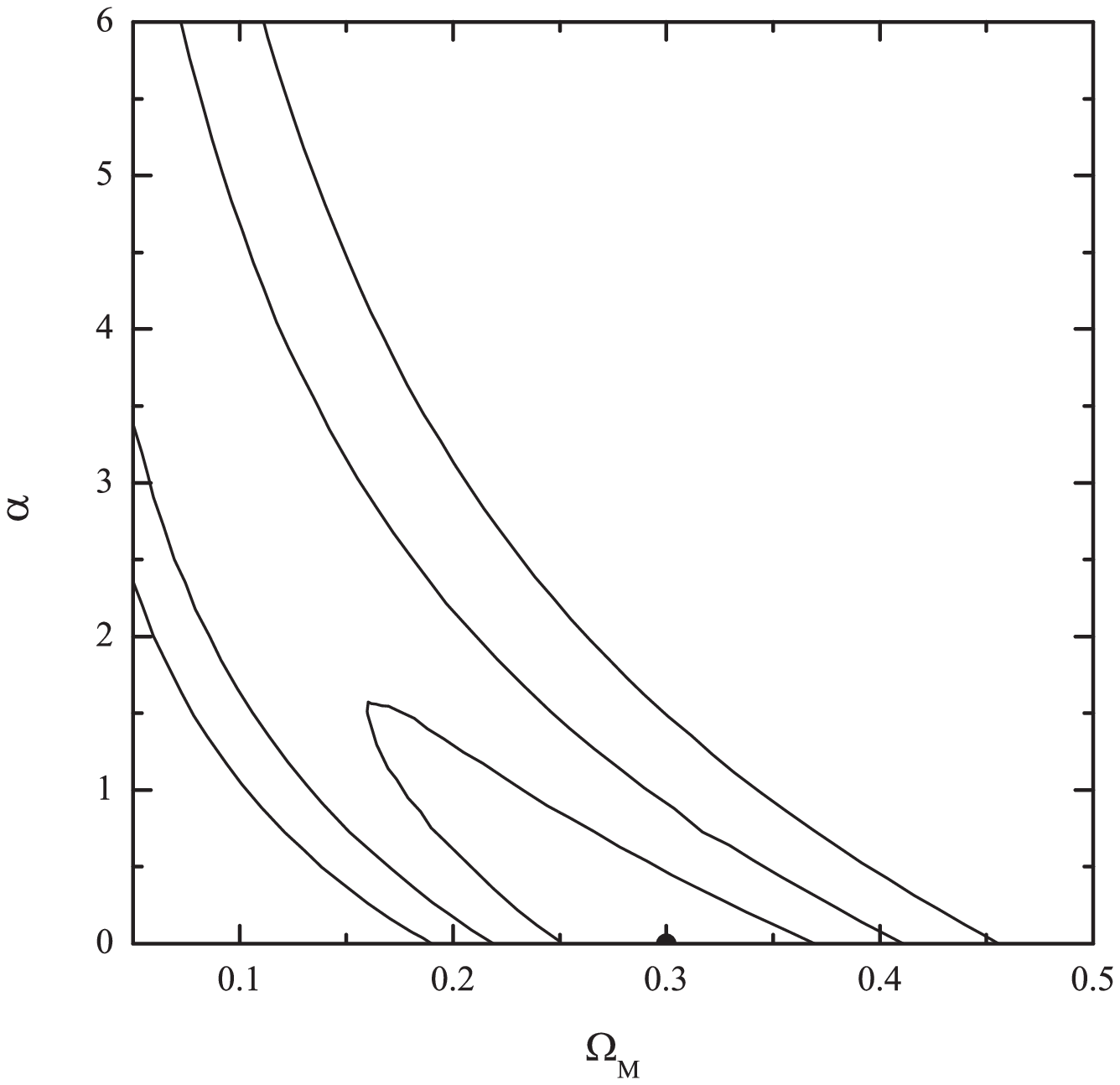}
\caption{Probability distribution function confidence contours (1, 2, and 3 $\sigma$ from inside to outside) for the $\phi$CDM model using
the R04 gold SNIa sample. The dot denotes the maximum likelihood and is at $\Omega_M = 0.30$ and $\alpha = 0.0$.}
\end{figure}

\begin{figure}[p] \psfig{file=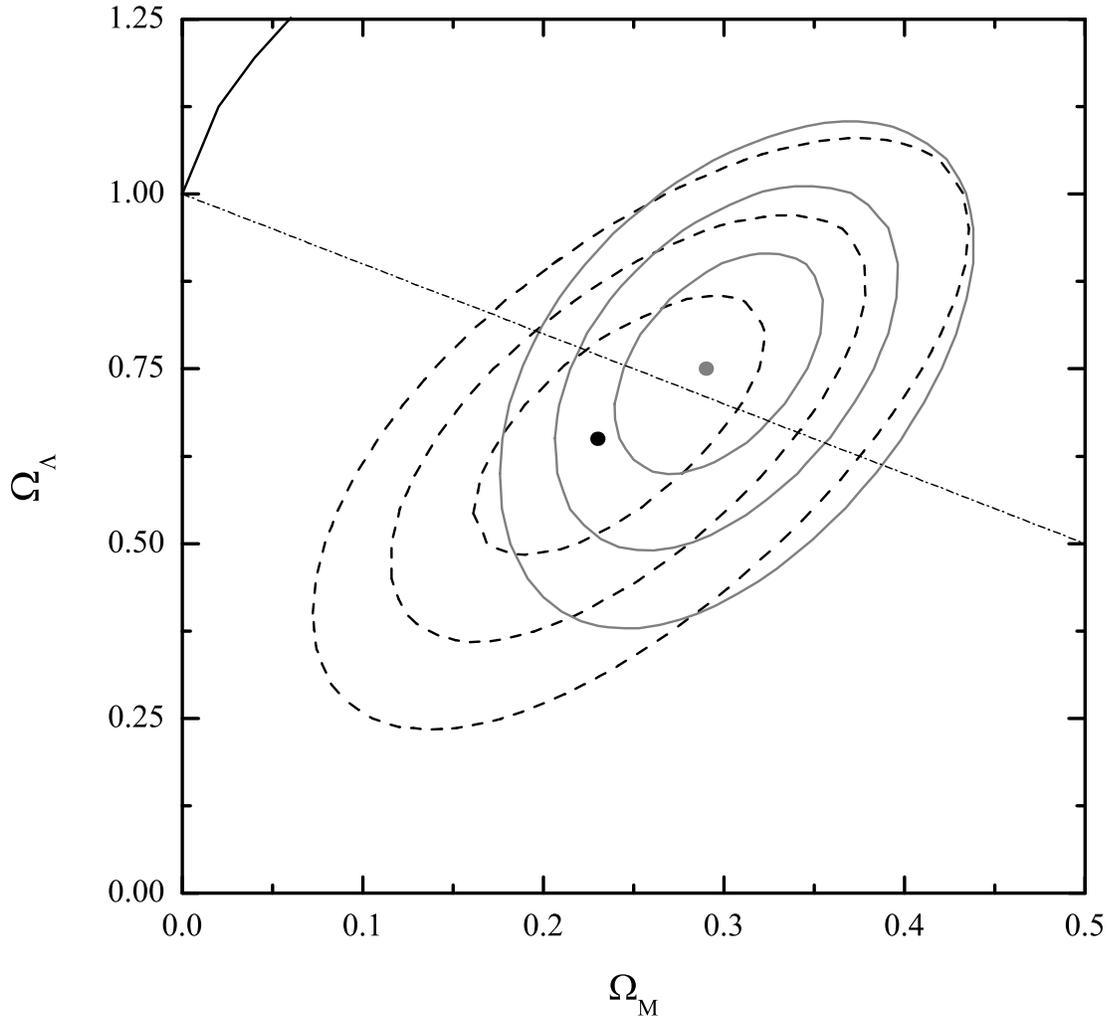}
\caption{Probability distribution function confidence contours (1, 2, and 3 $\sigma$ from inside to outside) for the $\Lambda$CDM model.
The solid gray lines are computed using $h = 0.72 \pm 0.08$ and $\Omega_b h^2 = 0.0214 \pm 0.002$, with maximum likelihood at $\Omega_M =
0.29$ and $\Omega_\Lambda = 0.75$. The black dotted lines use $h = 0.68 \pm 0.04$ and $\Omega_b h^2 = 0.014 \pm 0.004$, with maximum
likelihood at $\Omega_M = 0.23$ and $\Omega_\Lambda = 0.65$. The slanting dash-dot line indicates spatially-flat models and models to the
left and above the solid black line in the left top corner do not have a big bang.}
\end{figure}

\begin{figure}[p] \psfig{file=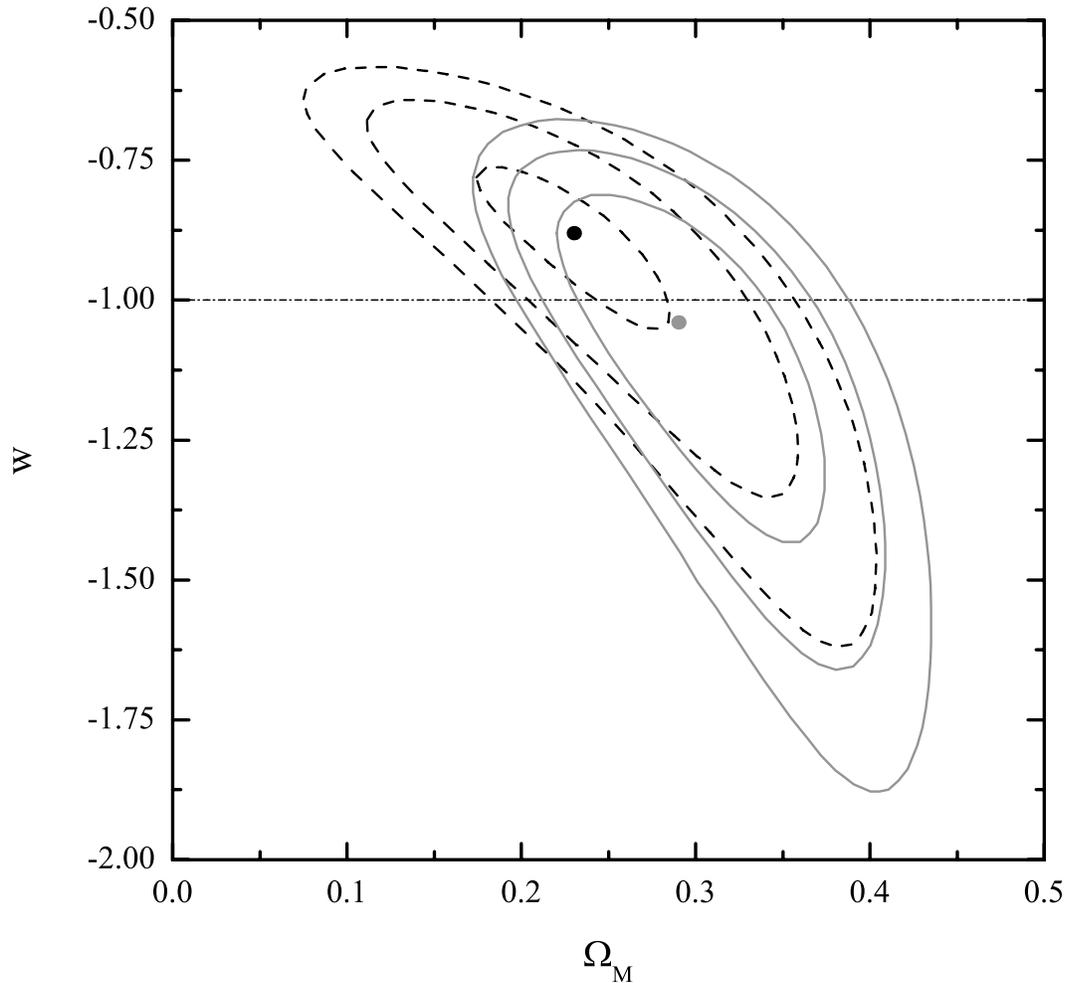}
\caption{Probability distribution function confidence contours (1, 2, and 3 $\sigma$ from inside to outside) for the XCDM parametrization.
The solid gray lines are computed using $h = 0.72 \pm 0.08$ and $\Omega_b h^2 = 0.0214 \pm 0.002$, with maximum likelihood at $\Omega_M =
0.29$ and $w = -1.04$. The black dotted lines use $h = 0.68 \pm 0.04$ and $\Omega_b h^2 = 0.014 \pm 0.004$, with maximum likelihood at
$\Omega_M = 0.23$ and $w = -0.88$. The horizontal dash-dot line at $w=-1$ indicates spatially-flat $\Lambda$CDM models.}
\end{figure}

\begin{figure}[p] \psfig{file=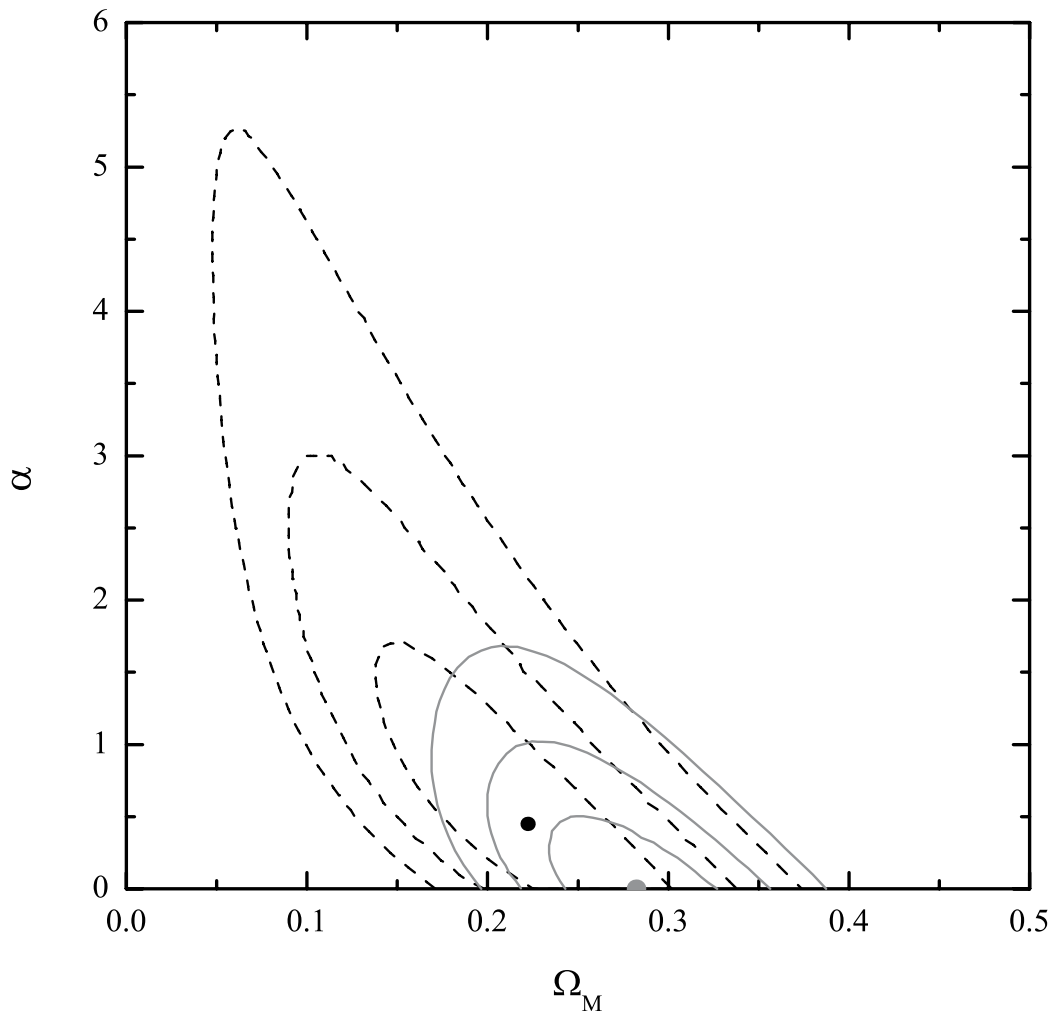}
\caption{Probability distribution function confidence contours (1, 2, and 3 $\sigma$ from inside to outside) for the $\phi$CDM model with $V(\phi)
\propto \phi^{-\alpha}$. The
solid gray lines are computed using $h = 0.72 \pm 0.08$ and $\Omega_b h^2 = 0.0214 \pm 0.002$, with maximum likelihood at $\Omega_M = 0.28$
and $\alpha = 0.0$. The black dotted lines use $h = 0.68 \pm 0.04$ and $\Omega_b h^2 = 0.014 \pm 0.004$, with maximum likelihood at
$\Omega_M = 0.22$ and $\alpha = 0.45$. The horizontal $\alpha = 0$ axis
corresponds to spatially-flat $\Lambda$CDM models.}
\end{figure}

\end{document}